\documentclass[prl,twocolumn,showpacs,preprintnumbers,amsmath,amssymb, floatfix]{revtex4}

\usepackage{graphicx}
\usepackage{dcolumn}
\usepackage{bm}

\usepackage{subfigure}

\begin{document}

\title{Quantum hacking: experimental demonstration of time-shift attack
against practical quantum key distribution systems}

\author{Yi Zhao, Chi-Hang Fred Fung, Bing Qi, Christine
Chen, Hoi-Kwong Lo}

\affiliation{Center for Quantum Information and Quantum Control,
Department of Physics and Department of Electrical \& Computer
Engineering, University of Toronto, Toronto, Ontario, M5S 3G4,
Canada.}

\date{\today}

\begin{abstract}
Quantum key distribution (QKD) systems can send signals over more
than 100 km standard optical fiber and are widely believed to be
secure. Here, we show experimentally for the first time a
technologically feasible attack, namely the time-shift attack,
against a commercial QKD system. Our result shows that, contrary to
popular belief, an eavesdropper, Eve, has a non-negligible
probability ($\sim$4\%) to break the security of the system. Eve's
success is due to the well-known detection efficiency loophole in
the experimental testing of Bell inequalities. Therefore, the
detection efficiency loophole plays a key role not only in
fundamental physics, but also in technological applications such as
QKD.
\end{abstract}

\maketitle

Quantum key distribution (QKD) \cite{BB84, Ekert91} provides a
method to share a secret key between legitimate users called
``Alice'' (the sender) and ``Bob'' (the receiver). The unconditional
security of QKD has ben rigorously proved based on the laws of
physics \cite{SecurityProofs,Deutsch:PRL96}. Even imperfect
practical QKD systems have also been proved secure assuming some
semi-realistic models \cite{GLLP, ILM}. The decoy method
\cite{DecoyTheory} was proposed to dramatically improve the
performance of a practical QKD system. Our group has implemented the
decoy method experimentally over 15km and 60km of telecom fibers
\cite{DecoyExperiment}. Incidentally, QKD has found real-life
applications in a recent Swiss election \cite{SwissElection}.

Recently, there has been a lot of theoretical interest on the
connection between the security of QKD and fundamental physical
principles such as the violation of Bell's inequality and the
no-signaling constraint \cite{Security:NoSignalingQKD} on space-like
observables. An ultimate goal, which has not yet been achieved
\cite{Security:DeviceIndependent}, is to construct a
device-independent security proof. As is well-known, the
experimental testing of Bell's inequality often suffers from the
detection efficiency loophole. A fair sampling assumption may save
the day. However, as we will demonstrate below, rather surprisingly,
the low detection efficiency of practical detectors not only
violates the fair sampling assumption that would be needed in
security proofs based on Bell-inequality violation, but also gives
Eve (an eavesdropper) a powerful handle to break the security of a
practical QKD system. Therefore, the detection efficiency loophole
is of both conceptual and practical interest.

Our work is an
illustration of general physical limitations, rather than a
particular technological weakness.
Indeed, a practical QKD system often includes two or more detectors. It
is virtually impossible to manufacture identical detectors in practice. As a result, the
two detectors of the same QKD system will exhibit different detection
efficiencies as functions of either one or a combination of
variables in the time, frequency, polarization, and/or spatial
domains. If Eve manipulates a signal in these variables, she could
effectively exploit the detection efficiency loophole to break the
security of a QKD system. In our experiment, we
consider Eve's manipulation of the time variable.
Our work demonstrates the general problem of detection efficiency loophole
in practical QKD systems.

Recently, quantum hacking has attracted much scientific and popular
attention \cite{News:TimeShift_Nature}. Makarov et al. proposed a
faked-state attack and studied its feasibility with their home-made
QKD system
\cite{ThrHack:Makarov_Mismatch,ThrHack:Makarov_FakedStatesSARG04}.
Unfortunately, this attack is an intercept-resend attack which is
hard to implement in practice. Therefore, this attack has never been
successfully demonstrated in experiments. Kim et al. simulated an
entanglement probe attack on the BB84 protocol
\cite{ExpHack:EntanglementProbe}. However, it serves to demonstrate
the security rather than the insecurity of QKD systems because this
attack has already been considered in standard security proofs. A
study of the information leakage due to public announcement of the
timing information by Bob was reported \cite{ExpHack:DetectionTime}.
However, Bob does not need to make such an announcement in practice.
In summary, despite numerous efforts, up till now,
no one has even come close to hacking successfully a practical QKD system, let alone a
commercial one.

Here, we present the first experimental demonstration of a
successful hacking against a commercial QKD system. It is highly
surprising to break a well-designed commercial QKD system with only
{\it current} technology. Our work shows clearly the slippery nature of
QKD \cite{GregorComments} and forces us to re-examine the
security of practical QKD systems and its applications in
real-life. The attack we use is the time-shift
attack proposed by us in \cite{ThrHack:TimeShift}. The time-shift attack
is simple to implement as it does not involve any measurement or
state
preparation by Eve. 

The time-shift attack exploits the detection efficiency mismatch
between the two detectors in a QKD system in the time domain. In QKD
security proofs (e.g. ref.\cite{GLLP}), a standard assumption is
that the detection efficiencies for the bits ``0'' and ``1'' are
equal. However, its validity is questionable
\cite{ThrHack:Makarov_Mismatch,ThrHack:Makarov_FakedStatesSARG04,ThrHack:TimeShift}.
For example, a typical time-dependence of the detection efficiency
of a practical fiber-based QKD system (with InGaAs avalanche photo
diodes (APDs) of telecom wavelength operating at gated Geiger mode)
is illustrated in Fig. \ref{Fig: Ideal Mismatch}. Note that, at time
$A$ the detection efficiency for the bit ``0'' is much higher than
that for the bit ``1'', while the opposite case can be found at $B$.
The detection efficiency mismatch can only be confidently removed if
the efficiencies are constant in time domain. We remark that even
non-gated single-photon detectors such as Si APDs exhibit detection
efficiency mismatch due to intrinsic dead-time
\cite{Security:Deadtime}.

The idea of the time-shift attack is simple.
Eve can shift the arrival time of each
signal to either $A$ or $B$ randomly with probabilities $p_A$ and
$p_B = 1- p_A$ respectively. Eve can carefully choose $p_A$ to keep
the number of
Bob's detection events of ``0''s and ``1''s equal. 
Since Bob's measurement result will be biased towards ``0'' or ``1''
depending on the time shift ($A$ or $B$), Eve can ``steal''
information without alerting Alice or Bob. A conceptual setup to
launch the time-shift attack is shown in Fig. \ref{Fig: Ideal
Setup}. Eve can choose to connect Alice and Bob through either a
longer arm or a shorter arm so as to shift the
signal 
around time $A$ (a negative
shift), 
or around time $B$ (a positive shift).




The success of our demonstration is a big surprise because  in our
experiment,  Eve {\it cannot} perform a quantum non-demolition (QND)
measurement on the photon number or compensate any loss introduced
by the attack, while Eve \emph{can} have arbitrarily advanced
technology in security proofs. In other words, our practical Eve is
much weaker than the eavesdropper in security proofs. It is
surprising to see an attack which can be implemented with current
technology (e.g., the time-shift attack) can do better than even the
QND attack, which is
significantly beyond current technology. 

%
The experiment is performed on top of a modified commercial ID-500 QKD setup
\cite{Stucki:NJP2002} manufactured by id Quantique. 
The schematic of our experimental setup is shown in Fig. \ref{Fig:
setup}. 

\begin{figure}\center
\subfigure[Conceptual efficiency mismatch.]{ \label{Fig: Ideal
Mismatch}
\includegraphics[width=3cm, height=1.3cm]{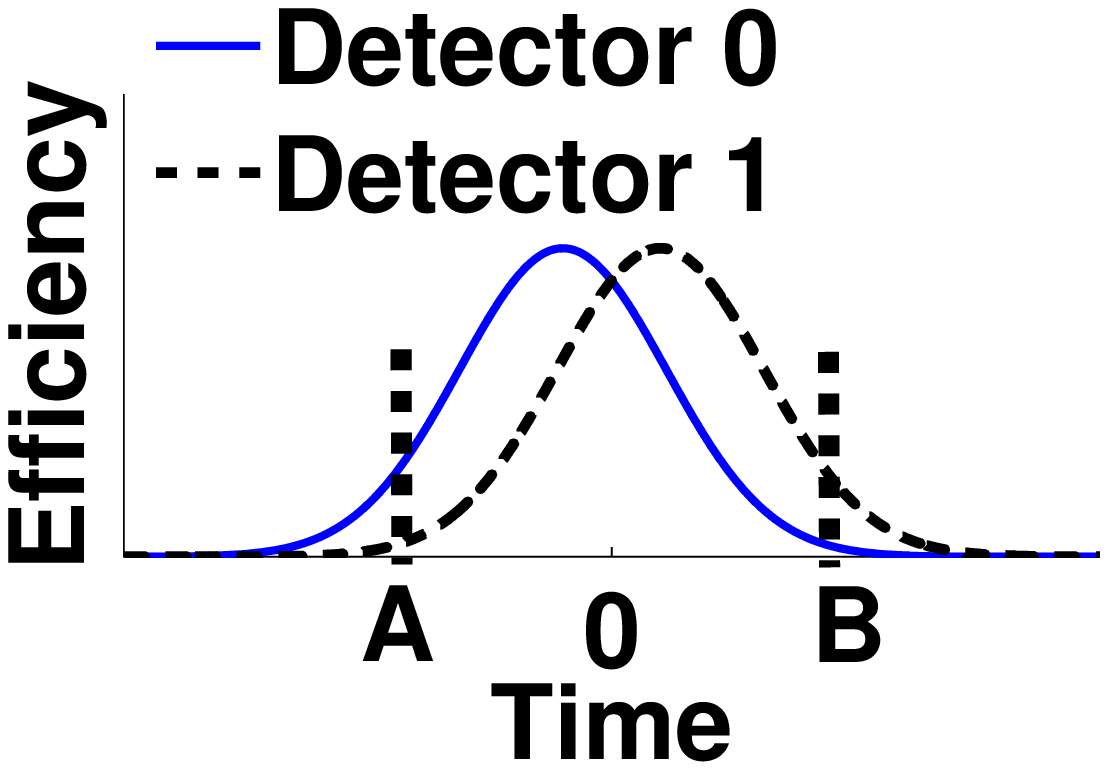}}
\subfigure[A conceptual schematic of Eve's attack. HOS: high-speed
optical switch.]{ \label{Fig: Ideal Setup}
\includegraphics[width=5cm]{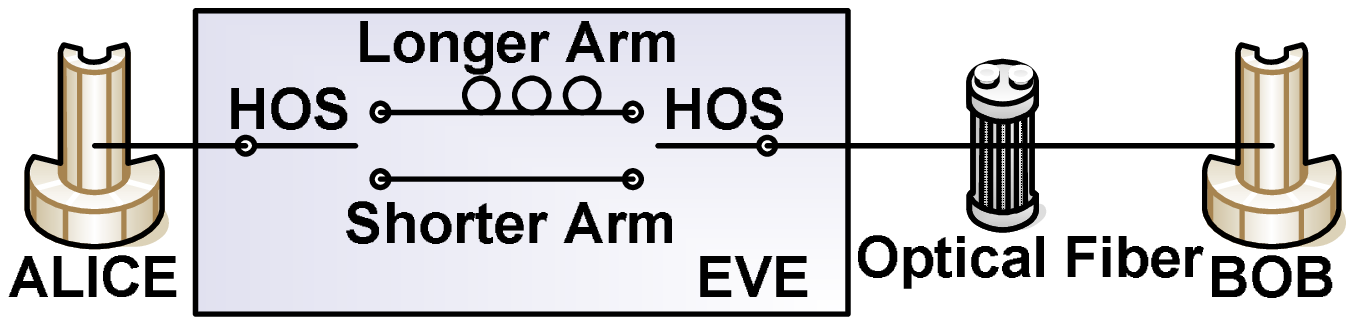}}\\
  \caption{Conceptual drawings.}\label{Fig: Conceptual}
\end{figure}

\begin{figure}\center
  \includegraphics[width=6cm]{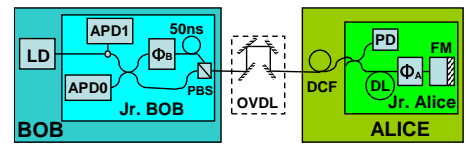}
  \caption{The schematic of experimental demonstration of the time-shift attack.
  Inside Jr. Bob/Jr. Alice: components in
Bob/Alice's package of id Quantique QKD system. Our modifications:
LD: narrow pulse laser diode; OVDL: optical variable delay line;
DCF: dispersion compensating fiber. Original QKD system: APD:
avalanche photon diode; $\Phi_\mathrm{A/B}$: phase modulator; PBS:
polarizing beam splitter; PD: classical photo detector; DL: delay
line; FM: faraday mirror. }\label{Fig: setup}
\end{figure}

The crucial issues in the experiment are the activation times of the
two detectors (APDs in Fig. \ref{Fig: setup}). The commercial QKD
system has a built-in calibration program which sets the activation
time of each detector independently. The activation times of the two
detectors differ slightly due to the discrepancies in the lengths of
the fibers connecting them. Ideally, to minimize the detection
efficiency mismatch, the difference of the activation times should
take a constant value. However, at times the difference in the
activation times as set by the built-in calibration program deviates
from this value, suggesting a larger efficiency mismatch. We
observed the maximum value of the deviations as  $\Delta\sim100$ps.
To get statistics of this deviation, we ran the built-in calibration
program for 2844 times, during which the deviation reaches $\Delta$
for 106 times. This is, the detection efficiency mismatch reaches
its maximum value with a probability of $\sim$4\%.



After the calibration of the activation times, we use the optical
variable delay line (
OVDL in Fig. \ref{Fig: setup}) to manually shift the arrival time of
the signals, looking for instants that show large efficiency
mismatch.

There are several challenges in this experiment. In our setup, the
gating window for the detectors (APDs in Fig. \ref{Fig: setup}) is
$\sim$500ps, which is close to the laser pulse width. This will
``blur'' the efficiency mismatch. However, the commercial QKD system
is not immune from the time-shift attack as Eve can simply apply
standard pulse compression technique to the bright pulses sent from
Bob to Alice in the channel (e.g.
\cite{Kibler:EL07}). 
In our experiment, we replaced the original laser source by a
PicoQuant laser diode (LD in Fig. \ref{Fig: setup}) with pulse width
 $\sim$100ps, which is equivalent to the compression scheme
mentioned above \cite{PulseCompression}.

Another challenge is the chromatic dispersion in the fiber which
broadens the laser pulses. We thus installed $\sim2$ km dispersion
compensating fiber 
(DCF in Fig. \ref{Fig: setup})
. Ideally, Eve can pre-chirp the bright pulses that are sent from
Bob to Alice. Note that both the pre-chirping and pulse compression
can be done on the bright pulses from Bob to Alice without touching
the quantum signal sent from Alice to Bob. Therefore, neither
process would increase the channel loss when Alice sends quantum
signals to Bob. We thus view the dispersion compensating fiber (DCF
in Fig. \ref{Fig: setup}) as part of Alice's local apparatus.

A third challenge is the optimization of the attack. Na\"{i}vely,
Eve could simply select large shifts as they would definitely
provide substantial intrinsic detection efficiency mismatches.
However, they may be suboptimal for the attack because their low
intrinsic detection efficiencies make the dark count significant,
increasing the quantum bit error rate (QBER) and consequently the
cost of the error correction. Therefore the task of choosing the
shifts is non-trivial. The time-shift attack will introduce
additional loss as the signals are shifted to the low-efficiency
region. Nonetheless, since Alice and Bob's channel may not be a
straight line and there may be additional loss due to components
such as optical switches, in practice Eve could lower the channel
loss by for example replacing existing channel with a better one
without alerting Alice and Bob. So, the power of time-shift attack
may be stronger than what one na\"{i}vely thinks.

We demonstrated the time-shift attack in the following way: first,
the activation times of detectors were determined by the built-in
program; second, the arrival times of the signals were shifted at a
step of 50 ps (a narrower step was not necessary as the pulse width
was $\sim$100 ps); third, at each shifted time, Alice and Bob
exchanged key at an average photon number (at Alice's output) of
0.1; fourth, Bob calculated the counts of each detector and the
error rates. The entire experiment after each calibration spanned
$\sim$15 minutes.
\begin{figure}
\center
  \includegraphics[width=7cm, height=3.5cm]{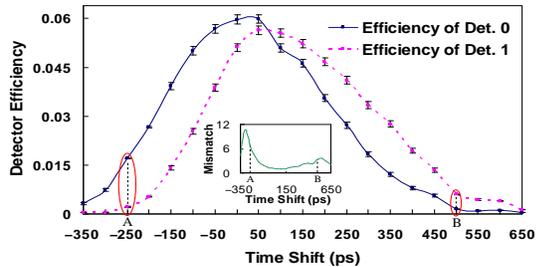}
 \caption{Efficiencies of the two detectors versus time shifts. Inset: the mismatch of detectors efficiencies
(defined
  as $\max(\frac{d_0}{d_1}, \frac{d_1}{d_0})$).  The peak
  efficiencies of detectors are slightly different, suggesting the detection efficiency has slightly drifted
  since the factory setting. The data size for time shifts with large detection efficiency
mismatch (-250ps, -200ps, 500ps, 600ps, and 650ps) is chosen to be
20.97Mbit to acquire accurate mismatch, while the data size for
other shifts is chosen to be 1.05Mbit to speed up the
experiment.}\label{Fig: ExpResult}
\end{figure}

In real attack Eve should apply an alternative technique to obtain
the efficiency mismatch as she has no access to Bob's apparatus
\cite{ThrHack:TimeShift}: she can gradually shift a small subset of
the signals and set them to 0 or 1 and deduce the amount of the
mismatch from Bob's detection announcement. Our experimental results
show that the mismatch is stable throughout the 15-min span of our
experiment. Therefore Eve has sufficient time to obtain the mismatch
information and launch her attack.

The experimentally measured detector efficiencies are shown in Fig.
\ref{Fig: ExpResult} for the case where the deviation in activation
times takes the maximal value $\Delta t_\mathrm{m}$. It shows
substantial detection efficiency mismatch. In particular, two shifts
with large mismatches
are found as in Table I. 

The security of the QKD system is analyzed in the following way: one
can estimate an upper bound $K_\text{U}$ of the key length
\emph{given} the efficiency mismatch known by Eve and  a lower bound
$K_\text{L}$ \emph{ignoring} the time-shift attack (as Alice and Bob
cannot detect the attack). If the upper bound is less than the lower
bound (i.e., $K_\text{L}>K_\text{U}$ ), there must be some
information leaked to Eve unknown to Alice or Bob.

\begin{table}[!b]\center
\caption{Experimental results.}\label{Tab: Experimental Result}
\subtable[The number of detections. 
]{\label{Tab: Raw Data}
\begin{tabular}{c c c c c c}
\hline Label & Shift (ps) & $d_0$ & $d_1$ & $N$\\
\hline A & -250 & 10992 & 1541 & 20,966,400 \\
\hline B & 500 & 1231 & 4059 & 20,966,400\\
\hline
\end{tabular}}
\linebreak \subtable[The number of detections given that Alice and
Bob use the same basis. $\tilde{N}=10,481,280$ bits. $Y$ is Bob's
bit value.]{\label{table-rawdata1}
\begin{tabular}{c c | c c}
\multicolumn{4}{c}{\small Time shift A ($-250$ps)}\\
\hline $Z_2$ & $X$ & $Y=1$ & $Y=0$ \\
\hline 0&1&336 & 139  \\
\hline 0&0&65 &  2557 \\
\hline 1&1&333&  120 \\
\hline 1&0&59&  2634\\
\hline 
\multicolumn{4}{l}{QBER: 0.06135} \\
\hline 
\end{tabular}
\linebreak \hspace{1cm}
\begin{tabular}{c c | c c}
\multicolumn{4}{c}{\small Time shift B ($500$ps)}\\
\hline $Z_2$ &$X$ & $Y=1$ & $Y=0$\\
\hline 0&1&979 & 31\\
\hline 0&0&41 &  260\\
\hline 1&1&1022&  37\\
\hline 1&0&35&  279\\
\hline 
\multicolumn{4}{l}{QBER: 0.05365} \\
\hline 
\end{tabular}}
\subtable[Parameters for computing the key length.] {\label{Tab:
Balanced}\label{table-lowerLoundParam}
\begin{tabular}{c c}
\multicolumn{2}{c}{\small Theoretical}\\
\hline $f(x)$ & $p_A$\\
\hline $1.22$ & $23.0$\%\\
\hline
\end{tabular}
\hspace{.1cm}
\begin{tabular}{c c c c c c}
\multicolumn{6}{c}{\small Experimental}\\
\hline $\mu$ &  $Y_0$ & $d_{0/1}$ & $E$ & $K_\mathrm{U}$ & $K_\mathrm{L}$\\
\hline $0.1$ &  $2.26 \times 10^{-5}$  & 3479 & 5.68\% & 1131bit & 1297bit \\
\hline
\end{tabular}
}
\end{table}

We consider that Alice sends $N$ bits to Bob, among which the same
basis are used for $\tilde{N}$ bits and Bob detects $\tilde{N} Q$
signals ($Q$ is the overall gain).
Here we assume that infinite decoy state protocol and one-way
classical communications for post-processing are used.

\emph{Lower bound}: The error correction will consume
\begin{eqnarray}
r_\text{EC}=\tilde{N} Q f(E) H_2(E) \label{eqn-ECC1}
\end{eqnarray}
bits, where $E$ is the overall QBER, $H_2(x)$ is the standard binary
Shannon entropy function, $f(x)$ is the
error correction inefficiency \cite{Brassard:ErrorCorrection}. The
net key length ignoring the time-shift attack is thus
\cite{GLLP,Lo:QIC05,Koashi2006,BenOr2005}
\begin{equation}\label{eqn-lower-bound-DR}
K_\text{L} = -r_\text{EC}+\tilde{N} \{ Q_1 [ 1- H_2(e_1) ] + Q_0\}
\end{equation}
where $Q_i$ and $e_i$ are the gain and the QBER for the signals with
$i$ photons sent by Alice.


\emph{Upper bound}: an upper bound
 is given
by \cite{Renner2005b}
\begin{eqnarray}\label{eqn-upper-bound-DR}\nonumber
K_\text{U} &=& -r_\text{EC}+\tilde{N}\cdot Q\cdot
\sum_{\substack{i=\{A,B\};j=\{0,1\}}}
[\text{Pr}\{Z_2=j|Z_1=i\}\\
&\cdot{}&\text{Pr}\{Z_1=i\}\cdot H_2(\mathrm{Pr}\{X=0|Z_1=i,
Z_2=j\})]
\end{eqnarray}
where $X$,  $Z_1$, and $Z_2$ are classical random variables
representing Alice's initial bit, Eve's choice of the time shift for
each bit, and the basis information, respectively.

The upper bound and the lower bound of the key rate can then be
calculated from Eqs.~\eqref{eqn-ECC1}-\eqref{eqn-upper-bound-DR}
using data in Table I. The calculation results are shown in Table
\ref{Tab: Balanced}. $Y_0$ is determined experimentally. Note that
no
double clicks were observed in our experiment.%
The fact that $K_\mathrm{L}>K_\mathrm{U}$ clearly indicates the
success of the attack \cite{SuccessDiscussion}.

We conclude with a few general lessons. First, counter-measures
often exist for known attacks. For instance, the ``four-state
measurement'' proposal (which suggests that for phase-encoding BB84
protocol, Bob's phase modulation is randomly
selected from a set of four values instead of two values) can shield the time-shift attack. 
Second, the implementation of a counter-measure may open up new
security loopholes. For instance, the four-state measurement scheme
will be vulnerable to combined large pulse
\cite{Gisin:TrojanHorse} and time-shift attack. 
Once an attack is known, the prevention is usually easy. However, we
have a third lesson: unanticipated attacks are most dangerous.

The time-shift attack is demonstrated on a bi-directional system.
However, it is a threat to a general class of QKD systems (including
uni-directional setup) and protocols (eg. \cite{Ekert91}). Moreover,
we are concerned with the general physical limitations of detection
efficiency loophole, rather than a specific technological problem.
The time-shift attack can be easily generalized to spatial-,
spectral-, and polarization-shift attack exploiting the efficiency
mismatch in the corresponding domains. On the practical side, our
work highlights the significance of side channel attacks
\cite{EQC,SECOQC:WhitePaper} in QKD. Historically, the existence of
a side-channel attack went back to the first QKD experiment, which
was unconditionally secure to any eavesdropper who happens to be
deaf!

In summary, we report the first experimental demonstration of a
technologically feasible attack against a commercial QKD system. Our
results clearly show that even QKD systems built by {\it
trustworthy} manufacturers may contain subtle flaws that will allow
Eve to break it with current technologies. The success of the attack
highlights the importance to battle-test practical QKD systems and
work on security proofs with {\it testable} assumptions. It is
remarkable that the detection efficiency loophole plays a key role
in both fundamental physics and technological applications (e.g.,
QKD systems) \cite{SECOQC:WhitePaper}. How to close the detection
efficiency loophole and side-channel attacks will be an important
subject for future investigation.






We thank generous help from A. Kurtsiefer, C. Lamas-Linares, X. Ma,
Vadim Makarov, and id Quantique. Support of the funding agencies
CFI, CIPI, the CRC program, CIFAR, MEDT, MITACS, NSERC, Perimeter
Institute, QuantumWorks, OIT, CQIQC, and PREA is gratefully
acknowledged. 

\bibliography{Z:/reference}

\begin{thebibliography}{31}
\bibitem[{Pul()}]{PulseCompression}
\bibinfo{note}{Pulse compression is a standard experimental technique that has
  been routinely performed in industries for decades. It would be of
  implementational interest to see the impact of the pulse compression on the
  time-shift attack, though not of physics interest.}
\end{thebibliography}

\bibliographystyle{apsrev}


\end{document}